\documentclass[
    ,final            
  ]
  {aipproc}

\layoutstyle{8x11double}

\begin{document}

\title{Two-phonon infrared processes in semiconductors}

\author{Hadley M. Lawler}{
  address={Department of Physics, University of Maryland, College Park, MD 20742}
}

\author{Eric L. Shirley}{
  address={Optical Technology Division, NIST, Gaithersburg, MD 20899-8441}
}

\begin{abstract}
 We report our detailed calculation of the infrared spectra of GaAs and GaP from first principles, and similar ongoing efforts for diamond-type spectra, which have recently been calculated.
 \end{abstract}

\maketitle


\section{Lattice Dynamics}
\indent Over the last decade or so, density-functional perturbation theory (DFPT) and application of the $2n+1$ theorem have revolutionized the theoretical
study of lattice dynamics~\cite{DFPT}.  Simultaneously, sophisticated frozen-phonon algorithms have been very successful in calculating complete phonon dispersion relations\cite{Kresse}.  These advances allow detailed knowledge of the vibrational continuum, to complement the realistic calculation of electronic band structure that has been possible for some time. 

\indent  Going beyond the harmonic approximation, with DFPT it has been possible to calculate anharmonic optical effects, such as widths, shapes, and shifts of Raman modes~\cite{Deb}, and two-phonon absorption spectra in silicon and germanium~\cite{Strauch}.  Higher-order anharmonic effects can be treated by combining DFPT and frozen-phonon methods, and developing both methods may prove important to widening the scope of multiphonon interactions and their effects which can be realistically studied from
theory.

\begin{figure}
  \includegraphics[height=.2\textheight]{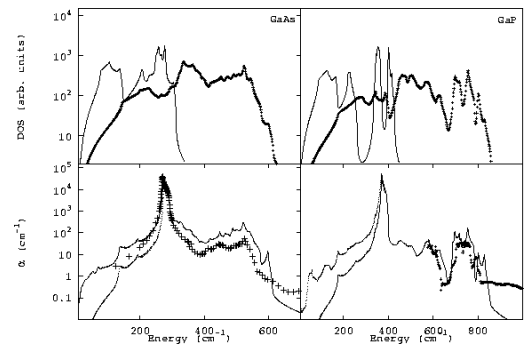}
  \caption{The one-phonon (line) and zero-momentum two-phonon ($+$) densities of states are plotted in top panels.  In bottom panels, the calculated (line) and measured (crosses) absorption coefficients are plotted at 300~K for GaAs, and <~10~K for GaP.  To point out the far-infrared temperature dependence, the theoretical spectra, in the far infrared, are plotted at low (GaAs) and room (GaP) temperatures (light dots).  Plotted theoretical spectral features have been moved to 0.98 of their calculated position to facilitate comparison with the experimental spectra.  The data are from Ref.~\cite{PalikAs}.}
\end{figure} 

\indent  Our focus has been on the frozen-phonon calculation of phonon-phonon and photon-phonon interactions.  We have calculated the absorption spectra for GaAs and GaP from the far infrared through twice the frequency of the dispersion oscillator, or the zero-momentum transverse-optical phonon.  The results demonstrate temperature dependence of the response in the far infrared, the asymmetric absorption profile of the dispersion oscillator and rich spectral structure caused by infrared light coupling to two-phonon states.  The correspondence with experimental spectra are shown in Figure~1.  We have also been able to relate prominent features in the two-phonon absorption spectra to very simple kinematical parameters of the primitive cell, the mass ratio of the basis ions in particular.  This work combines calculations of Raman widths--as has been accomplished with DFPT~\cite{Deb}--with the modern theory of polarization~\cite{DV} to calculate absorption spectra completely {\it{ab initio}}.

\indent  Currently we are working on a frozen two-phonon calculation of the weak infrared absorption in diamond-type materials.  Preliminary calculations of this nature have been reported~\cite{BP}, and at this conference, a thorough calculation of the spectra for Si and Ge with DFPT is presented~\cite{Strauch}.  A point worth emphasizing is that, with our two-phonon calculation, the adiabatic evolution gives no change in the ionic moment with respect to the supercell origin;  that is, the ionic contribution to the ``current'' is zero, because the the ionic displacements are given by standing waves.

\indent  All the work presented here is within the frozen-phonon approach.  In particular, the dynamical matrix and anharmonic matrix elements are calculated with wave-commensurate supercells, where a phonon's periodicity is mapped to a supercell~\cite{WCS}.  Born-von Karman force matrices can then be calculated as finite differences of Hellmann-Feynman forces.  The technique has the advantage that force matrices are sampled in reciprocal space, and then interpolated, as is also the case in applications of DFPT.  This allows implementation of the LO-TO splitting according to long-established formalisms~\cite{Gonze}.  It also avoids any aliasing.

\section{Zincblende Crystals}
\indent  Each of the above infrared-active processes is related to the interaction between zero-momentum two-phonon states and some macroscopic coordinate.  In the case of GaAs and GaP, within the present theory, it is only the dispersion oscillator which is directly coupled to the macroscopic field, and the two-phonon states acquire oscillator strength through their anharmonic coupling to the dispersion oscillator.  The relevant matrix element for this coupling is $M_{\alpha\beta{\bf k}}=\sum_{\tau\tau'ii'}
\frac{{\partial}{D^{\tau\tau'}_{ii'}({\bf k})}}{{\partial}u}
\epsilon^{{\bf k} \alpha}_{{\tau}i}\epsilon^{{-\bf k} \beta}_{{
\tau'}i'}$.  The derivative of the dynamical matrix, $D^{\tau\tau'}_{ii'}({\bf k})$, is taken with respect to the sublattice relative coordinate, and phonon polarizations are given on the right.  Phonon branches are denoted $\alpha$ and $\beta$, Cartesian directions $i$ and $i'$, and basis ions $\tau$ and $\tau'$.  The matrix element above is actually the cubic term in the Born-Oppenheimer potential which gives interaction between three normal modes of wave vectors ${\bf k},-{\bf k},$ and $0.$  

\indent  The two-phonon states, in addition to imprinting two-phonon character upon the spectra, also act as decay channels, giving the dispersion oscillator absorption feature finite linewidth and asymmetric shape.  Hence, the anharmonic theory of infrared spectra in polar materials accounts for a broad class of infrared optical properties, from sensitive temperature dependence of the far-infrared absorption, to the profile of the reststrahlen and the two-phonon features.  The dielectric function is given by\cite{diagrams}:
\begin{eqnarray}
\varepsilon_{ii'}({\omega}) -\varepsilon_{ii'}^{\infty} =
\nonumber
\frac{4\pi}{\Omega}
\sum_{{\nu=TO}}
\sum_{{\tau}{\tau}'jj'}
\frac{{Z^{\tau}_{ij}}{Z^{\tau'}_{i'j'}}\epsilon^{\nu}_{{\tau}j}\epsilon^{\nu}_{{\tau'}j'}}{2\sqrt{m_{\tau}m_{\tau'}}{\omega_{TO}}}\\
\left(
\frac{1}{\omega+\omega_{TO}+\Sigma_{TO}(\omega)}-\frac{1}{\omega-\omega_{TO}-\Sigma_{TO}(\omega)}
\label{ZB}
\right).
\end{eqnarray}
\nonumber
The numerator on the right side contains the Born effective charges, and the denominator contains the ionic masses.  The frequency of the dispersion oscillator and its self-energy appear in the quantity in parentheses.  The self-energy is temperature-dependent, and is calculated via diagrams~\cite{diagrams} and the matrix elements defined above.  The results for GaAs and GaP are plotted in Figure~1, along with the experimental spectra.

\section{Diamond-type Crystals}

\indent  Since the Born effective charges, which express the integrated oscillator strength in Eq.~\ref{ZB}, are zero in a diamond-type material, the absorption spectra follow from a different type of interaction.  In fact, the electric field at infrared frequencies is a macroscopic coordinate independent of the zero-momentum optical phonon coordinates, and the key quantity for the spectral calculation is, $\frac{{\partial}{D^{\tau\tau'}_{ii'}({\bf k})}}{{\partial}E}$, where $E$ is the electric field.  Such calculations were recently performed with DFPT and methods for incorporating macroscopic fields into electronic calculations~\cite{Strauch}.  

\indent  Another approach toward a calculation of the spectra is to evaluate the polarization change due to a two-phonon displacement: $\triangle{P}_{i}=\frac{\partial^2P_i}{\partial{u_{{\bf k}\alpha}}{\partial}{u_{-{\bf k}\beta}} }{u_{{\bf k}\alpha}}{u_{-{\bf k}\beta}}$, with the modern theory of polarization.  Initial work in this direction has been reported~\cite{BP}.  Using wave-commensurate supercells, ionic displacements represent standing waves, and are appropriate to the momentum-conserving two-phonon geometry.  For the diamond-type structure, there are six branches, and hence the polarization is a $6$~x~$6$ matrix at any particular wave vector, representing the possible combinations of phonon branches, making the desired quantity a third-rank tensor.  We are attempting to calculate this tensor in a Cartesian basis.

\IfFileExists{\jobname.bbl}{}
 {\typeout{}
  \typeout{******************************************}
  \typeout{** Please run "bibtex \jobname" to optain}
  \typeout{** the bibliography and then re-run LaTeX}
  \typeout{** twice to fix the references!}
  \typeout{******************************************}
  \typeout{}
 }

\end{document}